%% file: ms.tex
\documentclass[10pt,letterpaper]{proc}

\usepackage[top=0.6in, bottom=1.0in, left=0.6in, right=0.6in]{geometry}

\RequirePackage{calc}
\usepackage{times}
\usepackage{graphicx}
\usepackage{parskip}
\usepackage{epsfig}
\usepackage{wrapfig}
\usepackage{amsmath}
\usepackage{amssymb}
\usepackage{algorithm}
\usepackage{algorithmic}
\usepackage{float}
\usepackage{url}
\usepackage{color}

\floatname{algorithm}{Procedure}

\newcommand{\textunderscript}[1]{$_{\text{#1}}$}

\newcommand{\BO}     {\mathcal{O}}

\newcommand{\bv}[1]{\mathbf{#1}}           

\usepackage{cite}

\begin{document}
\title{Improved Shortest Path Maps with GPU Shaders}

\author{ \textbf{Renato Farias \hspace{1cm} Marcelo Kallmann} \\
Computer Science and Engineering Department\\
University of California, Merced,
CA, 95343 \\
E-mail: \{rfarias2,mkallmann\}@ucmerced.edu \\
}

\maketitle

\input{0-abstract}
\input{1-intro}

\input{2-rw}
\input{3-spms.tex}
\input{4-segments.tex}
\input{5-results.tex}
\input{6-discussion.tex}

\input{7-conclusions.tex}



\bibliographystyle{IEEEtran}

\bibliography{references}
%

\end{document}

%% file: 0-abstract.tex
\begin{abstract}

We present in this paper several improvements for computing shortest path maps using OpenGL shaders~\cite{Camporesi2014}. 
The approach explores GPU rasterization as a way to propagate optimal costs on a polygonal 2D environment, producing shortest path maps which can efficiently be queried at run-time.
Our improved method relies on Compute Shaders for improved performance, does not require any CPU pre-computation, and
handles shortest path maps both with source points and with line segment sources.
The produced path maps partition the input environment into regions sharing a same parent point along the shortest path to the closest source point or segment source.
Our method produces paths with global optimality, a characteristic which has been mostly neglected in animated virtual environments.
The proposed approach is particularly suitable for the animation of multiple agents moving toward the entrances or exits of a virtual environment, a situation which is efficiently represented with the proposed path maps. 

\end{abstract}

%% file: 1-intro.tex
\section{Introduction}
\label{sec:introduction}

\begin{figure}[ht]
  \centering
  \includegraphics[width=1\linewidth]{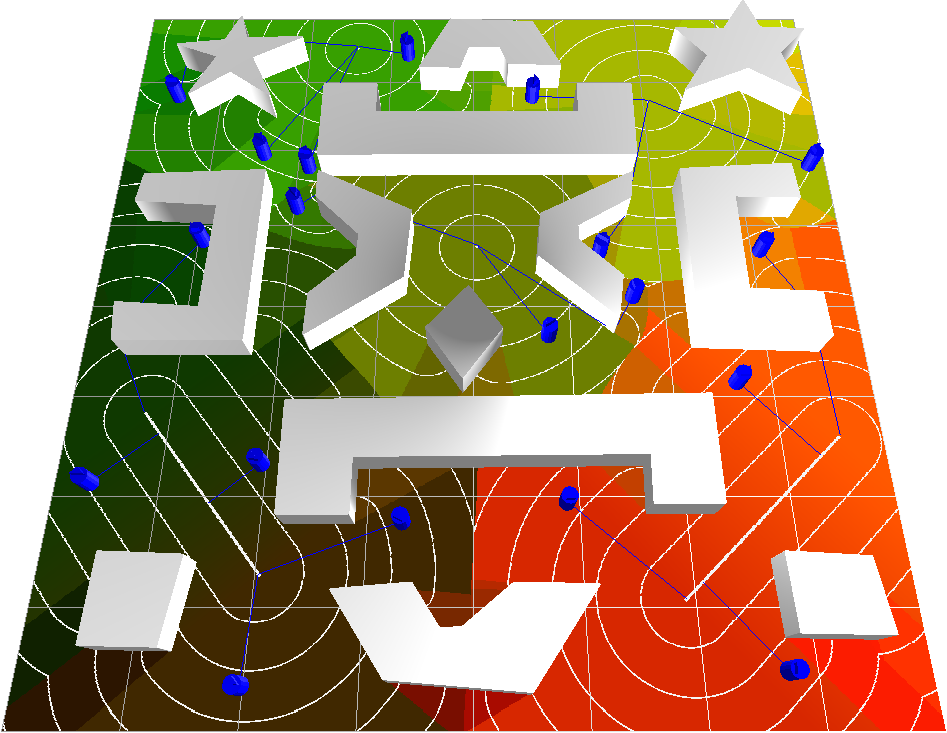}
  \caption{\label{fig:teaser}
  Example of a multi-source Shortest Path Map computed on a polygonal scene. There are three source points in the upper half of the scene, and two line segment sources in the lower half. The contour lines represent points with equal distances to their closest source. Contour lines are directly extracted from the distance field which is stored in the Z-Buffer as a result of our method. The blue cylinders are agents and each has a polygonal line representing its shortest path to the closest source.
           }
\end{figure}

Global navigation often depends on efficient path planning which is thus crucial in various applications from planning motions for real robots to controlling autonomous agents in virtual environments.
This paper focuses on the computation of optimal paths for agents in virtual environments. While several approaches have been introduced in recent years for computing paths among obstacles, the focus has mostly been on the efficiency of computation, without attention to providing global optimality guarantees.

This situation reflects the fact that computing optimal paths, or Euclidean shortest paths, efficiently is not a trivial task.
One way of computing Euclidean shortest paths is by constructing a visibility graph of the environment and then running graph search on it~\cite{DeBerg2008}. Unfortunately in the worst case the number of edges in the visibility graph is $\Theta(n^2)$, where $n$ is the number of vertices describing the obstacles, which can significantly slow down path queries based on search algorithms running on the graph.

Shortest Path Maps (SPMs) are constructed with respect to a ``source point'', and like Voronoi diagrams, SPMs partition the space into regions. Whereas regions in Voronoi diagrams share the same closest site, regions in SPMs share the same parent points along the shortest path to the source, which means that an SPM encodes shortest paths between a specified source and \textit{all} other points in a particular planar environment. 

While SPMs have been studied in Computational Geometry for several years, they have not been popular in practical applications. This is probably because their computation involves several complex steps, even  when considering non-optimal construction algorithms. The proposed GPU computation approach greatly simplifies the process of building SPMs, allowing them to be easily computed with rasterization procedures triggered from OpenGL shaders without any pre-computation.

Our approach introduces several advantages. 
While a CPU implementation requires some point localization technique in order to determine the region containing a query point, in the proposed GPU approach point localization is reduced to a simple constant time grid buffer mapping.
After this mapping, since every point in the SPM has direct access to its parent point along the shortest path to the closest source, agents have direct access to the next point to aim at when executing their trajectories. In addition, if the entire shortest path is needed it can be retrieved in linear time with respect to the number of vertices in the shortest path.

Our approach is based on the idea of cone rasterization from source points, line segment sources, and obstacle vertices. Unlike previous work of our group~\cite{Camporesi2014}, in the presented method we do not require pre-computation of the shortest path tree of the environment and we also do not need to create any geometry for the rasterized cones. Instead we use dedicated fragment shaders to simply fill in the pixels that have direct line-of-sight to the vertices, improving computation speed and also eliminating errors that were introduced from discretizing cone geometry into triangles.

Our shaders operate on the coordinates of the input vertices and when the buffer resolution is adequate our maps produce exact results not affected by the grid resolution.
Our approach is able to compute shortest path maps both with source points and with line segment sources, and can produce relatively complex dynamically-changing SPMs at real-time rates.

%% file: 2-rw.tex
\section{Related Work}
\label{sec:relatedwork}

Our work is related to different areas, from path planning and GPU computing to the computation of distance fields. The related work review below is organized according to these areas.

\textbf{Approaches to Path Planning}
Researchers in AI usually approach path planning with discrete search methods on grid-based environments, sometimes making use of hierarchical representations.
While several advancements on discrete search methods have been explored (heuristics, dynamic replanning, anytime planning, etc.), only a few attempts have focused on approximating Euclidean shortest paths~\cite{Nash2007}, and still not guaranteeing to achieve global optimality. 
A similar situation can be also observed 
in Computer Animation. While several approaches have been introduced in recent years, the state-of-the-art has focused mostly on the efficiency of computing collision-free paths  with the use of navigation meshes \cite{Geraerts10,Oliva13,Kallmann2014}, and has mostly neglected addressing global optimality.

One way to compute globally-optimal Euclidean shortest paths is to first build the visibility graph of the environment and then run a graph search algorithm on it \cite{Nilsson1969,DeBerg2008}. Previous work \cite{Lee84} has presented specific cases where the problem can be solved with greedy $\mathcal{O}(n\log{n})$ time algorithms without explicitly building the entire visibility graph. However, a visibility graph can have $\Theta(n^2)$ nodes, where $n$ is the number of vertices describing the environment, and a new graph search on it must be computed for each path query \cite{Welzl1985,Overmars1988,Storer1994}.
It is therefore difficult to develop efficient methods based on visibility graphs.


\textbf{Shortest Path Maps}
The first method related to 
Shortest Path Maps (SPMs) has worst-case time complexity  $\mathcal{O}(kn\log^2{n})$ \cite{Mitchell1991}, where $k$ is the ``illumination depth'', a parameter bounded above by the number of different obstacles touching a shortest path. Later, the first worst-case sub-quadratic algorithm was proposed applying the continuous Dijkstra expansion, which naturally leads to the construction of SPMs \cite{Mitchell1993}.
A nearly optimal algorithm for computing SPMs has been proposed taking optimal $\mathcal{O}(n\log{n})$ time to preprocess the environment, allowing distance-to-source queries to be answered in $\mathcal{O}(\log{n})$ time, and paths to be returned in $\mathcal{O}(\log{n+k})$ time, where $k$ is the number of turns along the path \cite{Hershberger99}.

Unfortunately, these methods and all the known algorithms with good theoretical running times involve complex techniques and data structures that overburden their practical implementation in applications.
In contrast, our GPU-based approach is relatively simple and has been able to produce SPMs of complexity not seen before in previous work.
Our benchmarks also demonstrate faster times in comparison to a previous GPU approach to compute SPMs~\cite{Wynters2013} (Table~\ref{tab:timecomparison}).


The idea of using shader rasterization as an efficient way to propagate wavefronts in the GPU was introduced by our group in 2014~\cite{Camporesi2014}. The method we present here significantly improves the approach in multiple ways:
1) we eliminate the need to precompute the visibility graph and SPT, and in doing so are able to easily address maps with multiple source points and line segment sources,
2) the speed of the method is improved with a new computation of shadow areas, and
3) we no longer need to construct actual geometry for the rendered cones simulating wavefront expansions; instead we simply employ a dedicated fragment shader to directly fill in the relevant pixels, simplifying the process and most importantly eliminating error accumulation from cone discretization.

\textbf{GPU Methods}
Previous work has investigated rasterization-based GPU techniques for related applications, in particular for computing Voronoi diagrams~\cite{Hoff1999}. 
Although we also employ rasterization techniques to accumulate distances,
our approach introduces the significant insight of placing primitives at accumulated heights in order to compute SPMs and represent optimal paths. 

GPU methods have also been explored for path planning from grid-based searches, for example by performing
multiple short-range searches in parallel~\cite{Henderson10},
by parallelizing expansions per-pixel on uniform grids~\cite{Kapadia13} and based on a quad-tree scheme~\cite{Kapadia14}.
However, grid-based approaches do not address global optimality in the Euclidean sense.
We nevertheless compare reported times from some of these works with our approach (Table \ref{tab:timecomparison}) and show that in addition to global optimality our method is also faster in most cases. 


\textbf{Distance Fields on Meshes}
Computing distance fields on meshes is a problem closely related to computing SPMs. The approach of Mitchell et al. \cite{Mitchell87} propagates front windows while solving front events during propagation, taking $\mathcal{O}(n^2 \log n)$ time. It is  possible to perform window propagation without handling all events~\cite{Chen1990,Xin2009}, reaching $\mathcal{O}(n^2)$ time but in practice processing a high amount of windows. Window prunning techniques have been investigated to improve practical running times~\cite{Qin2016}.
While a benchmark between our approach and these methods is left for future work, our approach represents a simpler solution for addressing the computation of distance fields and SPMs.

\textbf{Summary}
While several algorithms exist for computing globally-shortest paths and shortest path maps, available methods are either somewhat complex for practical use or too expensive for real-time applications.
The presented method is the first to be implemented entirely with GPU shaders, it does not require any pre-computation, and it enables multi-agent navigation based on paths with global optimality, a characteristic which has been neglected in simulated virtual environments developed to date.

%% file: 3-spms.tex
\section{Multi-Source Shortest Path Maps}
\label{sec:shortestpathmaps}

\begin{figure*}[htb]
  \centering
  \includegraphics[width=1\linewidth]{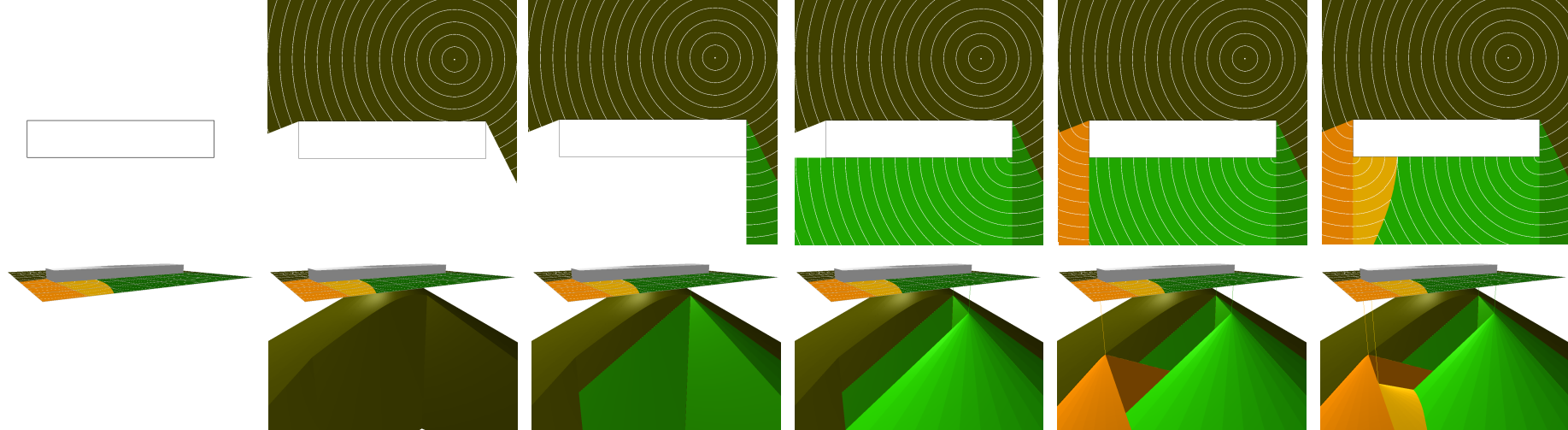}
  \caption{\label{fig:steps-ex}
Top row: steps for computing a single-source SPM in a simple scene.
Bottom row: corresponding 3D perspective view of each step.
           }
\end{figure*}

We first describe the base SPM case with multiple source points.
Let $n_s$ be source points $\{ \bv{s}_1, \bv{s}_2, ..., \bv{s}_{n_s} \}$ in the plane, 
such that $\bv{s}_i \in \mathcal{D}$, $i \in \{1,2,...,n_s\}$, 
and where $\mathcal{D} \subset {\mathbb{R}}$$^{2}$ defines a polygonal domain containing all sources.
In all our examples $\mathcal{D}$ is a rectangular area delimiting the environment of interest, and the GPU framebuffer will be configured to entirely cover $\mathcal{D}$. 
A set of polygonal obstacles $\mathcal{O}$, with a total of $n$ vertices, is also defined in $\mathcal{D}$ such that shortest paths will not cross any obstacles in $\mathcal{O}$.

Given source points the respective SPM will efficiently represent globally-shortest paths 
$\pi^*(\bv{p})$, which are optimal collision-free paths from any point $\bv{p} \in \mathcal{D}-\mathcal{O}$ to its closest source point $\bv{s}_i$
in the geodesic sense, i.e.,  
$\bv{s}_i$ is the source that minimizes 
$min_{i} \lambda^*(\bv{p},\bv{s}_i)=\lambda^*(\bv{p})$,
where $\lambda^*(\bv{p},\bv{s}_i)$ denotes the length of the shortest path $\pi^*(\bv{p},\bv{s}_i)$, $i \in \{1,2,...,n_s\}$.
Our SPM also efficiently represents the values of $\lambda^*$ for all pixels of the framebuffer by storing them in a dedicated buffer created in the OpenGL pipeline. This representation gives us direct access to the distance field of the environment and allows us to easily draw the white isolines that can be seen in most of the figures in this paper.
Depending on the situation source points can represent the start or the end point of a path. In most of the presented examples sources will represent goals to be reached by agents placed anywhere in the environment.

The plane represented by the framebuffer is located at $z = 0$. The basic idea of our method is to rasterize ``clipped cones'' with apices placed below source points and obstacle vertices, at $z$ heights equal to their $\lambda^*$ values, so that the final rendered result from an orthographic top-down view is the desired SPM (see Figure \ref{fig:steps-ex}).


The process is implemented as follows. An array containing the $n_s$ source points and $n$ obstacle vertices is stored in the GPU. At each iteration one point is copied into a reserved position of a data array where it will be used to rasterize a clipped cone. The point that is selected to generate the clipped cone at each iteration is referred to as that iteration's ``generator.'' Each point is processed once, such that the result is given after $n_s + n$ iterations.

Important to our approach is the fact that we do not actually need to create discretized geometry for representing and then drawing cones. Instead we simply fill in pixels that have direct line-of-sight to the generator, which is an equivalent operation. 
A cone apex is located below the generator relative to the $z=0$ plane. 
The depth values of the affected pixels increase proportionally to their Euclidean distances to the apex, as with the slope of a cone. Because the depth is accumulated over iterations, it represents the distance back to the source point along the shortest path, $\lambda^*$. When all clipped cones are drawn at their respective heights, the GPU's depth test will maintain, for each pixel, the correct parent generator point, which is the immediate next point on the shortest path from that pixel to the closest source point. We say that a cone ``loses'' to another at a given pixel when its depth is greater, leading it to be discarded in favor of the ``closer'' cone.

\subsection{Algorithm}

Given polygonal obstacles $\mathcal{O}$ with $n$ vertices and $n_s$ source points, $n_s \geq 1$, the total number of vertices to be processed is $n_{total} = n + n_s$. These points are stored in array {\scshape DataArray} of size $n_{total} + 1$.
The extra position is reserved for storing at each iteration the current generator that will be used for cone rasterization. By convention this is the first position in the array, {\scshape DataArray[0]}, and will be referred to as $\bv{g}_{cur}$. Once {\scshape DataArray} is constructed, it is stored in the GPU as a Shader Storage Buffer Object.
Each of the $n_{total} + 1$ positions in {\scshape DataArray} stores: \\ 
$\bullet$	$x,y$ : The original coordinates of the point in $\mathcal{D}$.\\
$\bullet$ {\scshape Status} : A flag that can be equal to {\scshape Source} for sources, {\scshape Obstacle} for obstacle vertices, or {\scshape Expanded} for points which have already generated a cone.\\
$\bullet$ {\scshape Distance} : The current known shortest path distance to the closest source point, $\lambda^*$. This will always be 0 for source points and is initially undetermined for obstacle vertices. \\
$\bullet$ {\scshape ParentId} : Array index into {\scshape DataArray} of the current parent point, which is the next point on the shortest path back to the closest source point. Since sources have no parent point, by convention they simply store their own index.

The framebuffer stores similar information for the pixels. For each pixel, its red and green components store the $x$ and $y$ coordinates of its parent point (equivalent to {\scshape DataArray[ParentId]}$.xy$), its blue component stores $\lambda^*$ (equivalent to {\scshape Distance}), and its alpha channel stores either $0$ if the pixel has yet to be reached by a cone or $>$0 otherwise. When the buffer is drawn, the color of each pixel is mapped in the following way: $x$ is used as the red component, $y$ is used as the green component, and the blue component is zeroed. Although this mapping is arbitrary, it allows to visualize the location of a region's parent from the red and green intensities. 

The SPM generation consists of four steps which repeat $n_{total}$ times such that each point is processed once. 
The steps are presented in Procedures 1-4. The hat notation (e.g., $\bv{\widehat{n}}$) denotes unit vectors.

Step 1 is a search in {\scshape DataArray} where the position with the smallest {\scshape Distance} is copied into the reserved position of the array, index 0.
Only points which have not yet generated a cone ({\scshape Status} $\neq$ {\scshape Expanded}) are considered in this search, and once a point is chosen its status is updated to {\scshape Expanded} so that it cannot be processed again.
The point that is chosen becomes $\bv{g}_{cur}$, the current generator. This step can be skipped in the first iteration of the algorithm as we can just start with one of the source points. 

\begin{algorithm}
\caption{Search Compute Shader}
\label{alg:shader-search}
\begin{algorithmic}[1]

  \REQUIRE {\scshape DataArray}

  \STATE int $generatorId \leftarrow -1$
  \STATE float $generatorDist \leftarrow -1$
	
	\FOR{$\forall i, i \in {1, 2, ..., n_{total}}$}
	  \IF{{\scshape DataArray[}$i${\scshape].Status} $\neq$ {\scshape Expanded}}
	    \IF{{\scshape DataArray[}$i${\scshape].Status} = {\scshape Source} \OR ($generatorId = -1$ \OR {\scshape DataArray[}$i${\scshape].Distance} $< generatorDist$)}
			  \STATE $generatorId$ $\leftarrow$ $i$
	      \STATE $generatorDist$ $\leftarrow$ {\scshape DataArray[}$i${\scshape].Distance}
		  \ENDIF
		\ENDIF
	\ENDFOR
	
	\STATE {\scshape DataArray[0]} $\leftarrow$ {\scshape DataArray[}$generatorId${\scshape]}
	\STATE {\scshape DataArray[}$generatorId${\scshape].Status} $\leftarrow$ {\scshape Expanded}

\end{algorithmic}
\end{algorithm}

\begin{figure}[htb]
  \centering
  \includegraphics[width=1\linewidth]{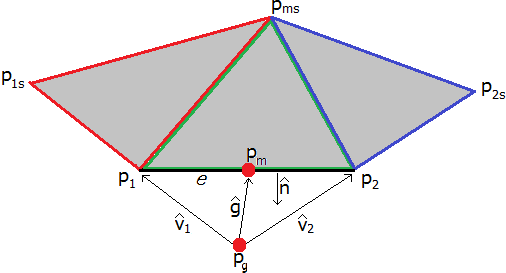}
  \caption{\label{fig:shadowarea}
  Example of a shadow area. The line segment \textit{e} represents the side of an obstacle. The red point \textbf{p\textunderscript{g}} is the generator, points \textbf{p\textunderscript{1}} and \textbf{p\textunderscript{2}} are the endpoints of \textit{e}, and point \textbf{p\textunderscript{m}} is the middle point of \textit{e}. Vectors $\bv{\widehat{v}}_1$, $\bv{\widehat{v}}_2$, and $\bv{\widehat{g}}$ are the normalized vectors from \textbf{p\textunderscript{g}} to \textbf{p\textunderscript{1}}, \textbf{p\textunderscript{g}} to \textbf{p\textunderscript{2}}, and \textbf{p\textunderscript{g}} to \textbf{p\textunderscript{m}}, respectively. Points \textbf{p\textunderscript{1s}}, \textbf{p\textunderscript{2s}}, and \textbf{p\textunderscript{ms}} are calculated in the following way: \textbf{p\textunderscript{1s}} = \textbf{p\textunderscript{1}} $+$ $c_{svf}\bv{\widehat{v}}_1$, \textbf{p\textunderscript{2s}} = \textbf{p\textunderscript{2}} $+$ $c_{svf}\bv{\widehat{v}}_2$, and \textbf{p\textunderscript{ms}} = \textbf{p\textunderscript{m}} $+$ $c_{svf}\bv{\widehat{g}}$. The three triangles are sufficient to cover the entire area behind the segment. Using less than three triangles may not result in a correct shadow if the generator is close to the segment because the area becomes wide and thin. Value $4$ is used for constant $c_{svf}$ 
such that shadows of any size can be handled given that our obstacle coordinates are normalized.
           }
\end{figure}

Step 2 is to generate a shadow area in order to solve visibility constraints. Using a geometry shader, we draw into a stencil buffer three triangles behind every obstacle line segment that is front-facing with respect to $\bv{g}_{cur}$, in a manner illustrated in Figure \ref{fig:shadowarea}. Any pixel covered by one of these triangles is considered to be in shadow. The resulting buffer is used as a stencil buffer in the next step. Three triangles is the minimum number of triangles needed to cover all possible shadow shapes. We use constant $c_{svf}>0$, which stands for shadow vector factor, when computing the points that make up the triangles. This constant must be large enough to handle shadows of all sizes. Since our coordinates are OpenGL normalized coordinates in the $\left[-1,1\right]$ range, a value of 4 is always enough.

\begin{algorithm}
\caption{Shadow Area Geometry Shader}
\label{alg:shader-shadow}
\begin{algorithmic}[1]

  \REQUIRE {\scshape DataArray}
	\REQUIRE $\bv{g_{cur}}$ \COMMENT{Current generator point}
	\REQUIRE \textit{e} \COMMENT{One of the sides of a scene obstacle}

  \STATE vec4 \textbf{p\textunderscript{1}} $\leftarrow$ first endpoint of \textit{e}
  \STATE vec4 \textbf{p\textunderscript{2}} $\leftarrow$ second endpoint of \textit{e}
	\STATE vec4 \textbf{p\textunderscript{m}} $\leftarrow$ (\textbf{p\textunderscript{1}}$+$\textbf{p\textunderscript{2}})/$2$
	\STATE vec4 \textbf{p\textunderscript{g}} $\leftarrow$ project and normalize vec4( $\bv{g_{cur}}.xy$, 0, 0 )
	
  \STATE float $dx \leftarrow$ \textbf{p\textunderscript{2}}$.x$ $-$ \textbf{p\textunderscript{1}}$.x$
	\STATE float $dy \leftarrow$ \textbf{p\textunderscript{2}}$.y$ $-$ \textbf{p\textunderscript{1}}$.y$

	\STATE vec4 $\bv{\widehat{g}}$ $\leftarrow$ normalize( \textbf{p\textunderscript{m}} $-$ \textbf{p\textunderscript{g}} )
	\STATE vec4 $\bv{\widehat{n}}$ $\leftarrow$ normalize( vec4( $dy, -dx, 0, 0$ ) )

	\STATE float $d \leftarrow$ dot( $\bv{\widehat{g}}$, $\bv{\widehat{n}}$ )

	\IF{$d < 0.1$}
		\STATE vec4 $\bv{\widehat{v}}_1$ $\leftarrow$ normalize( \textbf{p\textunderscript{1}} $-$ \textbf{p\textunderscript{g}} )
		\STATE vec4 $\bv{\widehat{v}}_2$ $\leftarrow$ normalize( \textbf{p\textunderscript{2}} $-$ \textbf{p\textunderscript{g}} )

		\STATE vec4 \textbf{p\textunderscript{1s}} $\leftarrow$ \textbf{p\textunderscript{1}} $+$ $c_{svf}\bv{\widehat{v}}_1$
		\STATE vec4 \textbf{p\textunderscript{2s}} $\leftarrow$ \textbf{p\textunderscript{2}} $+$ $c_{svf}\bv{\widehat{v}}_2$
		\STATE vec4 \textbf{p\textunderscript{ms}} $\leftarrow$ \textbf{p\textunderscript{m}} $+$ $c_{svf}\bv{\widehat{g}}$

	  \STATE EmitPrimitive( \textbf{p\textunderscript{1}}, \textbf{p\textunderscript{ms}}, \textbf{p\textunderscript{1s}} )
	  \STATE EmitPrimitive( \textbf{p\textunderscript{2}}, \textbf{p\textunderscript{2s}}, \textbf{p\textunderscript{ms}} )
	  \STATE EmitPrimitive( \textbf{p\textunderscript{1}}, \textbf{p\textunderscript{2}}, \textbf{p\textunderscript{ms}} )
	\ENDIF

\end{algorithmic}
\end{algorithm}

Step 3 draws a clipped cone with the generator $\bv{g}_{cur}$ directly above its apex along the $z$ axis. As previously stated, we do not actually create geometry for the cone but instead simply run a fragment shader over every pixel on the screen. The pixels that are not in shadow have direct line-of-sight to $\bv{g}_{cur}$, so they calculate their Euclidean distance to $\bv{g}_{cur}$ and add it to $\bv{g}_{cur}$'s accumulated distance, {\scshape Distance}. If this sum is smaller than the current {\scshape Distance} of the pixel (from the cone of a previous $\bv{g}_{cur}$), then its {\scshape Distance} is updated and its {\scshape ParentId} is set to $\bv{g}_{cur}$'s index.

Finally, step 4 is to update the {\scshape Distance} of all points visible from the current generator, in a way similar to step 3. Each point not in shadow calculates its distance to $\bv{g}_{cur}$ plus $\bv{g}_{cur}$'s {\scshape Distance}, and if that sum is smaller than its previous {\scshape Distance} it stores the new {\scshape Distance} and $\bv{g}_{cur}$'s index in its {\scshape ParentId}. The reason steps 3 and 4 are separate is because step 3 is updating the framebuffer, while step 4 is updating the {\scshape DataArray}.

After all points have been processed, which means $n_{total}$ iterations of steps 1-4, the result in the framebuffer will be the desired SPM.
Examples of SPMs with a single source point are shown in Figure \ref{fig:results1} and with multiple source points are shown in Figure \ref{fig:results2}.

The search in step 1 is $\BO(n_{total})$ because it is a sequential search in the array of size $n_{total}$.
Step 2 is $\BO(r)$, where $r$ is the resolution of the framebuffer, because in the worst case there will be enough triangles to render every pixel in the buffer. 
Step 3 is likewise $\BO(r)$.
Step 4 takes $\BO(n)$ time because potentially every obstacle vertex can have its distance updated. 
Steps 2, 3, and 4 are however executed in parallel by the GPU. Step 1 leads to a quadratic overall algorithm because the steps are iterated $n_{total}$ times; however, we have not observed any need to optimize this step as in our experiments this step represented about 1\% of the total runtime cost.

\begin{algorithm}
\caption{Cone Fragment Shader}
\label{alg:shader-cone}
\begin{algorithmic}[1]

  \REQUIRE {\scshape DataArray}
	\REQUIRE $\bv{g_{cur}}$ \COMMENT{Current generator point}
	\REQUIRE $fragCoord$ \COMMENT{$xy$ coordinates of the pixel}
	\ENSURE vec4 $fragValue$

	\STATE bool $inShadow \leftarrow$ is the pixel in shadow or not?
	\STATE vec4 $currentValue \leftarrow$ what's currently stored in this pixel \COMMENT{Texture fetch}
  \STATE vec4 $fragValue \leftarrow currentValue$ \COMMENT{If nothing else, pass the current value on}

	\IF{$inShadow$ = false}
	  \STATE vec2 \textbf{p} $\leftarrow$ normalize $fragCoord$
	  \STATE vec2 \textbf{p\textunderscript{g}} $\leftarrow$ project and normalize $\bv{g_{cur}}.xy$
    \STATE float $newDist \leftarrow$ distance( \textbf{p}, \textbf{p\textunderscript{g}} ) $+$ $\bv{g_{cur}}$.{\scshape Distance}
		
	  \IF{there is no currently stored distance in the pixel \OR $newDist <$ $currentValue.z$}
	    \STATE $fragValue \leftarrow$ vec4( $\bv{g_{cur}}$$.xy, newDist, 1$ )
	  \ENDIF
	\ENDIF

\end{algorithmic}
\end{algorithm}

\begin{algorithm}
\caption{Distance Compute Shader}
\label{alg:shader-compute}
\begin{algorithmic}[1]

  \REQUIRE {\scshape DataArray}
	\REQUIRE $\bv{g_{cur}}$ \COMMENT{Current generator point}
	
	\STATE int $id \leftarrow$ index of the point to be updated
	\STATE bool $inShadow \leftarrow$ is the point in shadow or not?

  \IF{$inShadow$ = false}
	  \STATE vec2 \textbf{p} $\leftarrow$ project and normalize {\scshape DataArray}$[id].xy$
	  \STATE vec2 \textbf{p\textunderscript{g}} $\leftarrow$ project and normalize $\bv{g_{cur}}.xy$
    \STATE float $newDist \leftarrow$ distance( \textbf{p}, \textbf{p\textunderscript{g}} ) $+$ $\bv{g_{cur}}$.{\scshape Distance}
	
		\IF{there is no currently stored distance in {\scshape DataArray}$[id]$ \OR $newDist <$ {\scshape DataArray}$[id].${\scshape Distance}}
	    \STATE {\scshape DataArray}$[id].${\scshape Distance} $\leftarrow newDist$
			\STATE {\scshape DataArray}$[id].${\scshape ParentId} $\leftarrow$ $\bv{g_{cur}}$'s original index
	  \ENDIF
	\ENDIF

\end{algorithmic}
\end{algorithm}

%% file: 4-segments.tex
\section{Segment Sources}
\label{sec:segmentsources}

Line segment sources are one natural extension to our method, and are interesting as sources for what they can represent. Many goals in real-world scenarios are not single points but line segments, for example the finish line of a race, the thresholds of doorways or hallways, and the boundary of a coastline can all be represented as polygonal line segments. For instance, many of these cases appear when planning evacuation routes from buildings. Being able to compute SPMs with segments as sources allows us to maintain global optimality in these practical situations.

Consider that we now have additional $n_l$ line segment sources $\{ \bv{l}_1, \bv{l}_2, ..., \bv{l}_{n_l} \}$,
such that $\bv{l}_i$, $i \in \{1,2,...,n_l\}$,
consists of two endpoints $\in \mathcal{D} - \mathcal{O}$.
The SPM will then efficiently represent globally-shortest paths 
$\pi^*(\bv{p})$, which are now optimal collision-free paths from any point $\bv{p} \in \mathcal{D}-\mathcal{O}$ to the closest reachable point on its closest segment source $\bv{l}_i$, in the geodesic sense.

\begin{figure}[bh]
  \centering
  \includegraphics[width=0.49\linewidth]{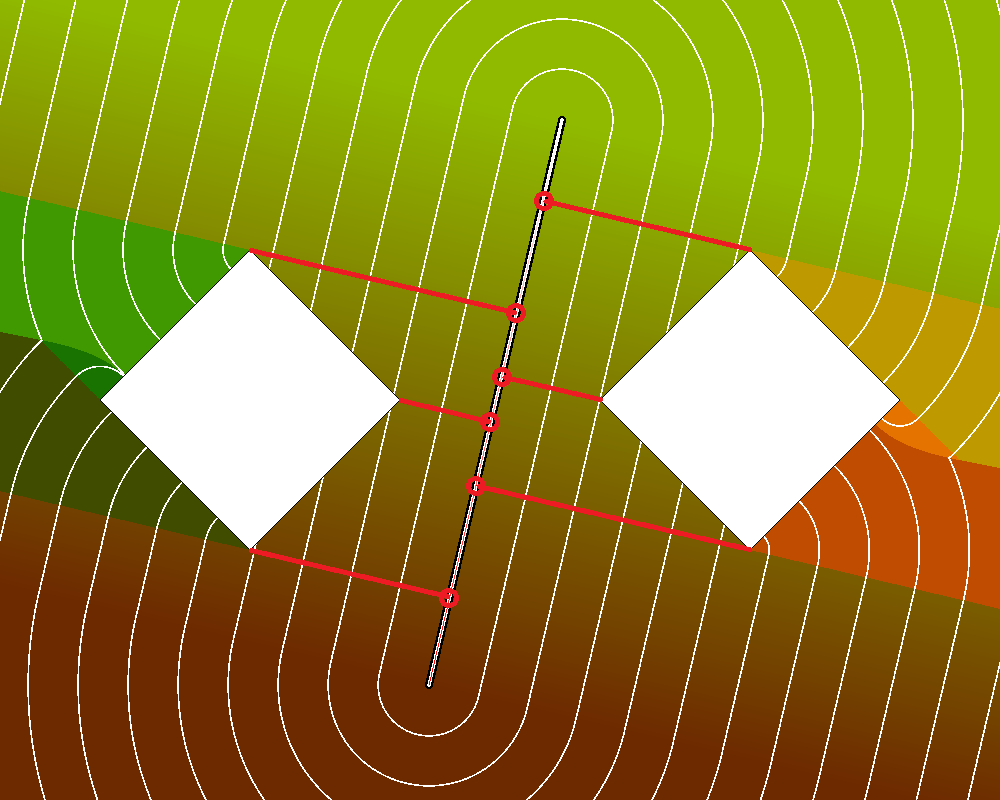}
  \includegraphics[width=0.49\linewidth]{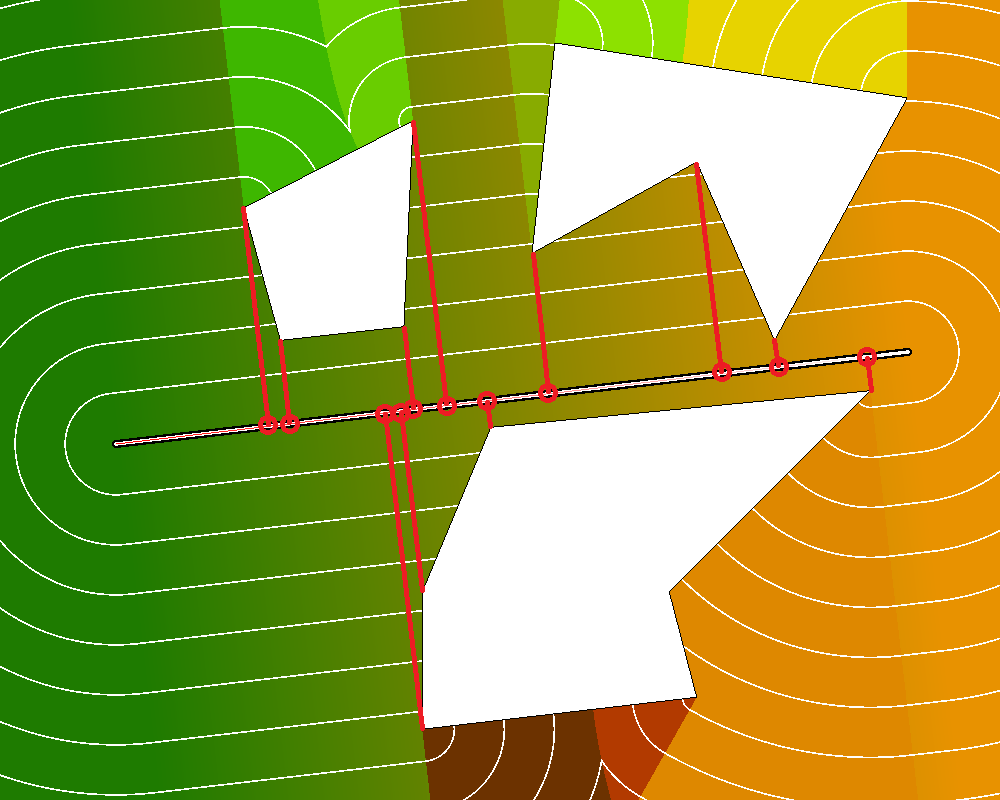}
  \caption{\label{fig:critpoints-ex}
  The circled points on the segment sources are the critical points, which are projections of obstacle vertices.
  }
\end{figure}

Every line segment $\bv{l}_i$ can have $n_{c_i}$ \textit{critical points}, $n_{c_i} \geq 0$. A critical point denotes a point on the segment onto which at least one obstacle vertex projects. The obstacle vertex must have direct line-of-sight to the segment. Critical points are where the visibility of the scene changes with respect to the segment and are useful because in practice every path that passes through the corresponding obstacle vertex will have its shortest path reach the line segment on that critical point (see Figure \ref{fig:critpoints-ex}).
For each $\bv{l}_i$, first the two endpoints of the segment create two entries in {\scshape DataArray} which are treated identically to source points. Then, $n_{c_i} + 1$ further entries are created, where $n_{c_i}$ is equal to the number of critical points segment $\bv{l}_i$ possesses.
Every one of these entries stores two pairs of $xy$ coordinates rather than just one, with {\scshape Status} set to {\scshape SourceSegment}, to represent the sub-segments of $\bv{l}_i$.
If $n_{c_i} = 0$, then the two endpoints are simply used because the segment has no sub-segments. If $n_{c_i} > 0$, then every adjacent pair of points, including both endpoints and critical points, will create an entry in {\scshape DataArray}.

The distance calculation of the SPM generation process is different when the generator's {\scshape Status} is marked as {\scshape SourceSegment}. It is necessary to determine whether the point being updated is closer to one of the endpoints of the sub-segment, or somewhere inbetween. If it is closer to one of the endpoints, the distance is simply the distance to that endpoint. Otherwise, the distance is equal to the distance between the point and its projection on the sub-segment.

\begin{figure}[bh]
  \centering
  \includegraphics[width=0.49\linewidth]{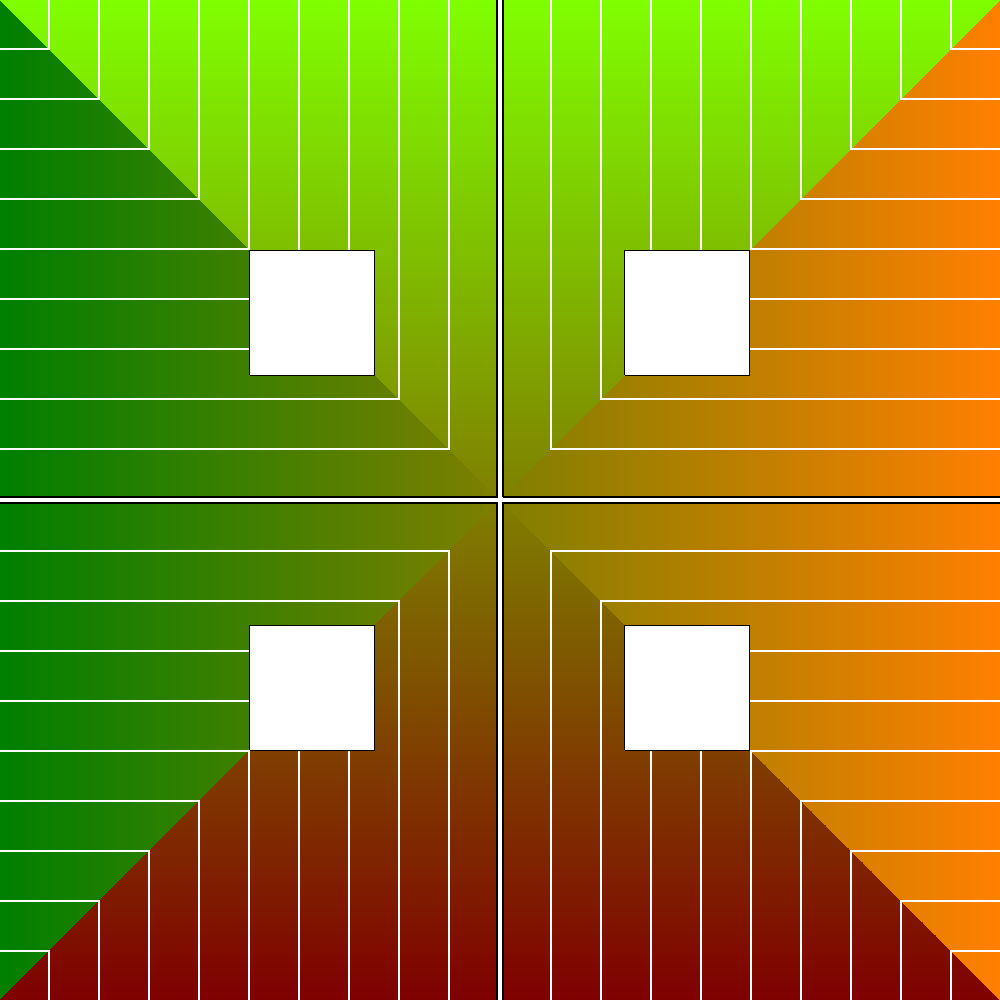}
  \includegraphics[width=0.49\linewidth]{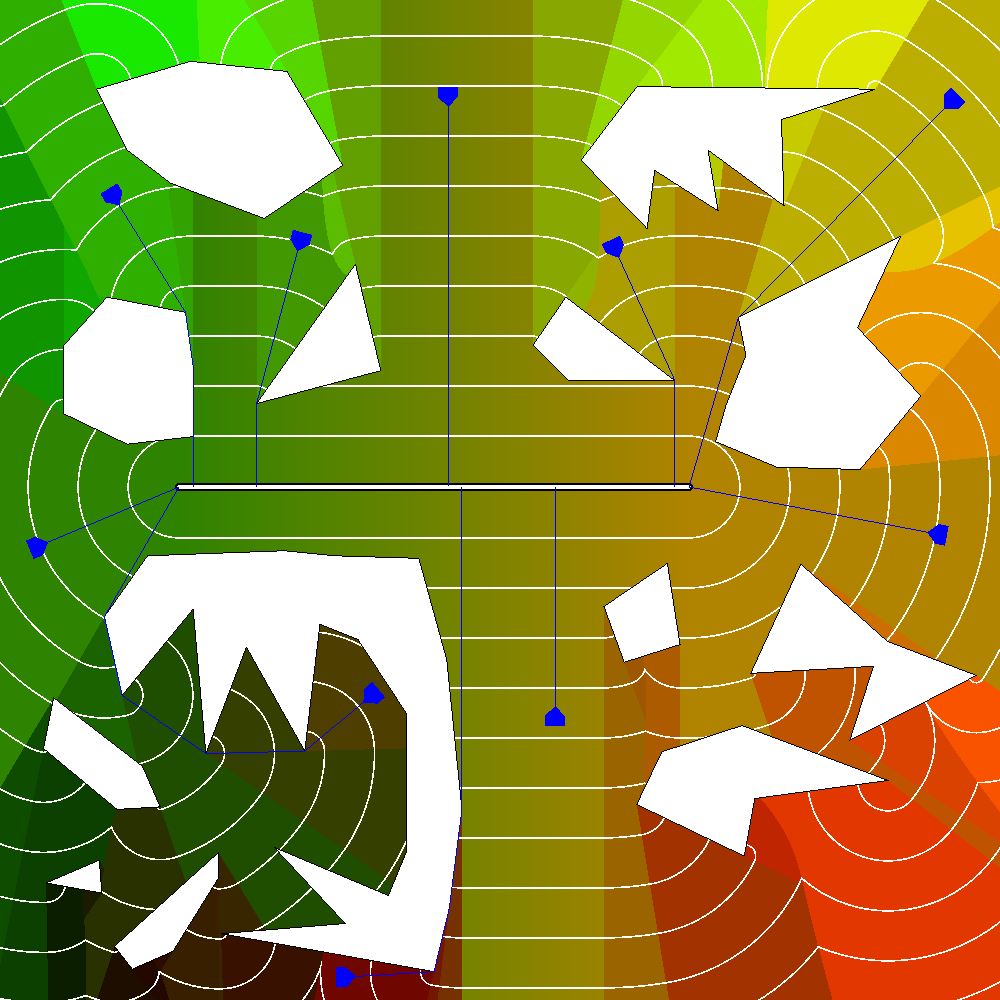}
  \caption{\label{fig:lineseg-ex}
Line segment source examples.
Left: SPM of two segment sources intersecting at the center.
Right: Several paths from agents represented as blue triangles to their closest points in a segment source. 
In both cases the white contours represent the distance field from the sources.
           }
\end{figure}

The described changes are sufficient to handle both points and line segments as sources. Figure \ref{fig:lineseg-ex} shows additional examples of SPMs with line segment sources. 

%% file: 5-results.tex
\section{Results and Discussion}
\label{sec:results}

\input{5-resultsTabs.tex}


We evaluate the performance of our method with several benchmarks
using a framebuffer resolution of 1000x1000 on a Nvidia GeForce GTX 970 GPU and an Intel Core i7 3.40 GHz computer with 16GB of memory.

Table \ref{tab:spmtime} shows average execution times for computing 100 single-source SPMs with random source points in $\mathcal{D}-\mathcal{O}$. The table shows times both with and without transferring the resulting SPM back to the host memory.

Figure \ref{fig:spmtimechart} charts out computation times on the Profiling maps.
These maps are composed of uniform rows of square obstacles (in the same layout as the rightmost environment in Figure \ref{fig:results1})
with large visible areas from all points in the map. 
This represent a worst-case scenario for our method because the amount of points that have to be considered at each step is almost the maximum. Still we observe that the increase in computation time is close to linear.


Table \ref{tab:timecomparison} shows that our method is also able to compute SPMs and return optimal paths faster than some previous GPU-based methods which are grid-based and non-optimal. For example, Kapadia et al. \cite{Kapadia13} gives times to plan paths on a grid environment with similar resolution to the buffer used in our benchmarks, 1024x1024, as follows: between 32.931 and 49.126 seconds for a GT 650M and between 21.246 and 30.778 seconds for a GTX 680. While our benchmarks used a newer GTX 970 GPU, we nevertheless believe that a new card would not offer the significant speed up to match even the 2.80 second running time we achieved on our most complicated map. In a later work a quad-tree was employed to significantly speed up the computation \cite{Kapadia14}, but sacrificing  optimality even more in the process. 


\input{5-resultsFigs.tex}

%% file: 5-resultsTabs.tex
\begin{table}[htbp]
	\centering
		\begin{tabular}{*{5}{|c}|}
			\cline{4-5} \multicolumn{3}{c}{} & \multicolumn{1}{|c|}{\bf Comp.} & \multicolumn{1}{|c|}{\bf Comp.+Transfer} \\ 
			\hline
			\bf Map name & \bf P & \bf V & \bf Time (s) & \bf Time (s) \\
			\hline
			\bf Simple1 & 3 & 13 & 0.0011 & 0.0209 \\
			\hline
			\bf Simple2 & 4 & 16 & 0.0018 & 0.0221 \\
			\hline
			\bf Concave1 & 2 & 12 & 0.0011 & 0.0207 \\
			\hline
			\bf Concave2 & 13 & 96 & 0.0088 & 0.0465 \\
			\hline
			\bf Spiral & 1 & 38 & 0.0022 & 0.0274 \\
			\hline
			\bf SpmEx1 & 3 & 15 & 0.0016 & 0.0215 \\
			\hline
			\bf SpmEx2 & 13 & 91 & 0.0100 & 0.0470 \\
			\hline
			\hline
			\bf Profiling0 & 4 & 16 & 0.0014 & 0.0221 \\
			\hline
			\bf Profiling1 & 16 & 64 & 0.0054 & 0.0404 \\
			\hline
			\bf Profiling2 & 36 & 144 & 0.0456 & 0.0858 \\
			\hline
			\bf Profiling3 & 64 & 256 & 0.1251 & 0.1680 \\
			\hline
			\bf Profiling4 & 100 & 400 & 0.2701 & 0.3099 \\
			\hline
			\bf Profiling5 & 196 & 784 & 0.8564 & 0.8863 \\
			\hline
			\bf Profiling6 & 400 & 1600 & 2.7371 & 2.8070 \\
			\hline
		\end{tabular}
	\caption{Average time in seconds to compute a single-source SPM on various maps (shown in \ref{fig:results1}). P and V are the number of polygons and vertices. 
  }
	\label{tab:spmtime}
\end{table}

\begin{figure}[htb]
  \centering
  \includegraphics[width=0.99\linewidth]{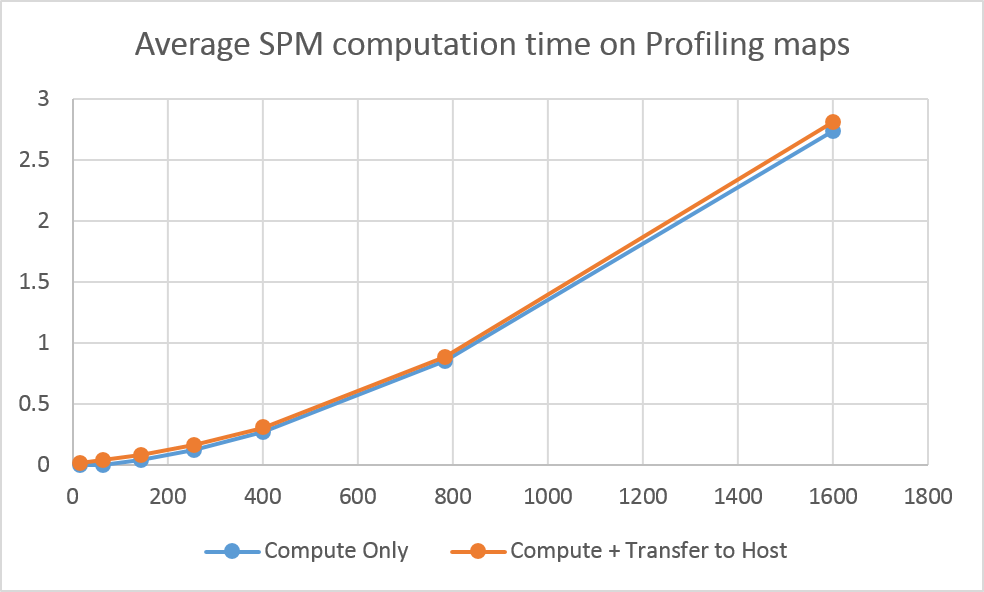}
  \caption
  {\label{fig:spmtimechart}
  The $x$ axis represents the number of obstacle vertices in the scene, and the $y$ axis represents the computation time in seconds.
  }
\end{figure}

\begin{table*}[htbp]
	\centering
		\begin{tabular}{*{8}{|c}|}
			\hline
			\bf Method & \bf CPU & \bf GPU & \bf Resolution & \bf O & \bf V & \bf Optimality & \bf Time (s) \\
			\hline
			\bf Dynamic Search using uniform grid (1) & -- & GF GT 650M & 1024x1024 & -- & -- & Average & 32.93 \\
			\hline
			\bf Dynamic Search using uniform grid (1) & -- & GF GTX 680 & 1024x1024 & -- & -- & Average & 21.25 \\
			\hline
			\bf Dynamic Search using uniform grid (2) & -- & -- & 1024x1024 & -- & -- & Average & 14.12 \\
			\hline
			\bf Dynamic Search using quad-tree (2) & -- & -- & 1024x1024 & -- & -- & No & 0.04 \\
			\hline
			\bf CUDA-based SPM (3) & i7 2.66 GHz & GF GTX 580 & 1024x1024 & 64 & 256 & Best & 1.42 \\
			\hline
			\bf Previous Shader-based SPM (4) & i7 3.40 GHz & GF GTX 570 & 1024x1024 & 64 & 256 & Best & ~0.11-0.17 \\
			\hline
			\bf Current Shader-based SPM (5) & i7 3.40 GHz & GF GTX 970 & 1000x1000 & 64 & 256 & Best & 0.13 \\
			\hline
		\end{tabular}
	\caption{A comparison of GPU-based techniques: (1) Dynamic search using an uniform grid [Kapadia et al. 2013]; (2) Dynamic search using an uniform grid or a quad-tree [Garcia et al. 2014]; (3) [Wynters 2013]; (4) [Camporesi and Kallmann 2014]; (5) Our method. The resolution refers to the resolution of the grid or the framebuffer, depending on the method. The number of obstacles (O) and vertices (V) in the environment are included for the last three methods because they affect running times. Some hardware details were not specified in the papers. 
	}
	\label{tab:timecomparison}
\end{table*}

%% file: 5-resultsFigs.tex
\begin{figure*}[htb]
  \centering
  \includegraphics[width=0.24\linewidth]{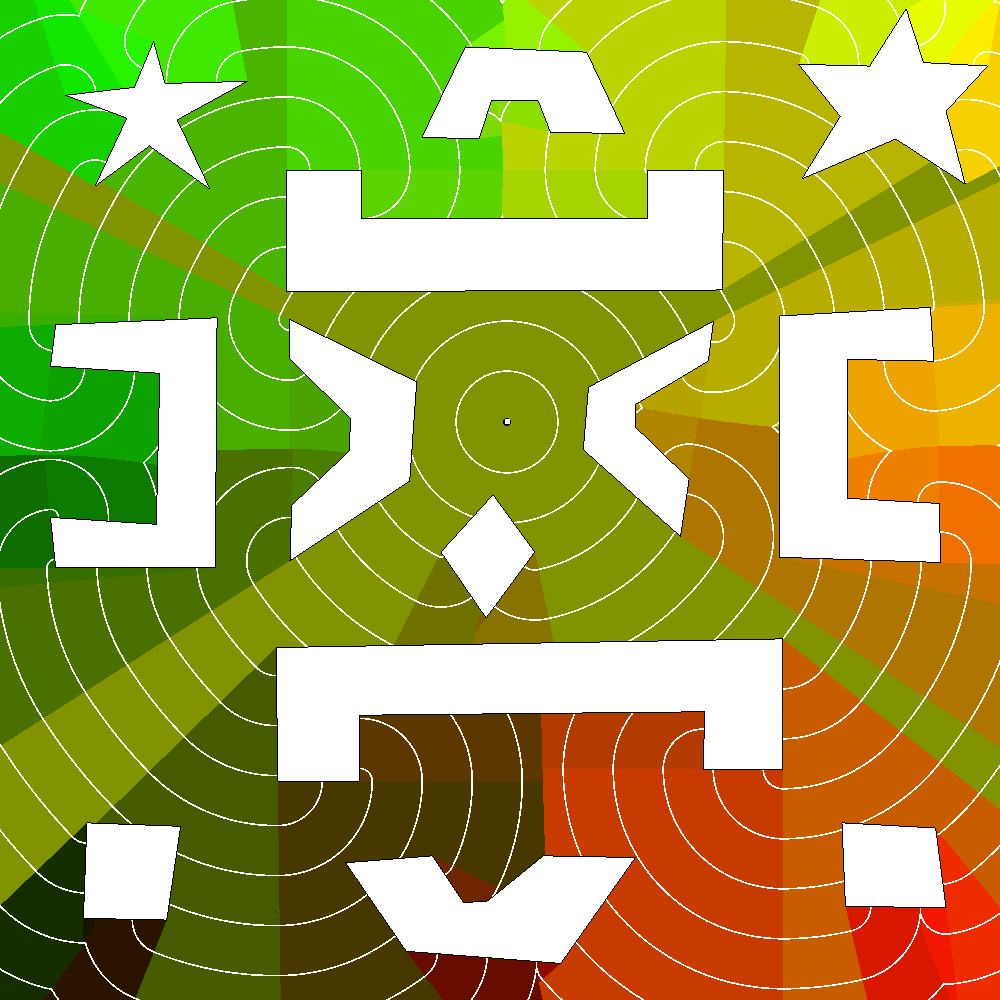}
  \includegraphics[width=0.24\linewidth]{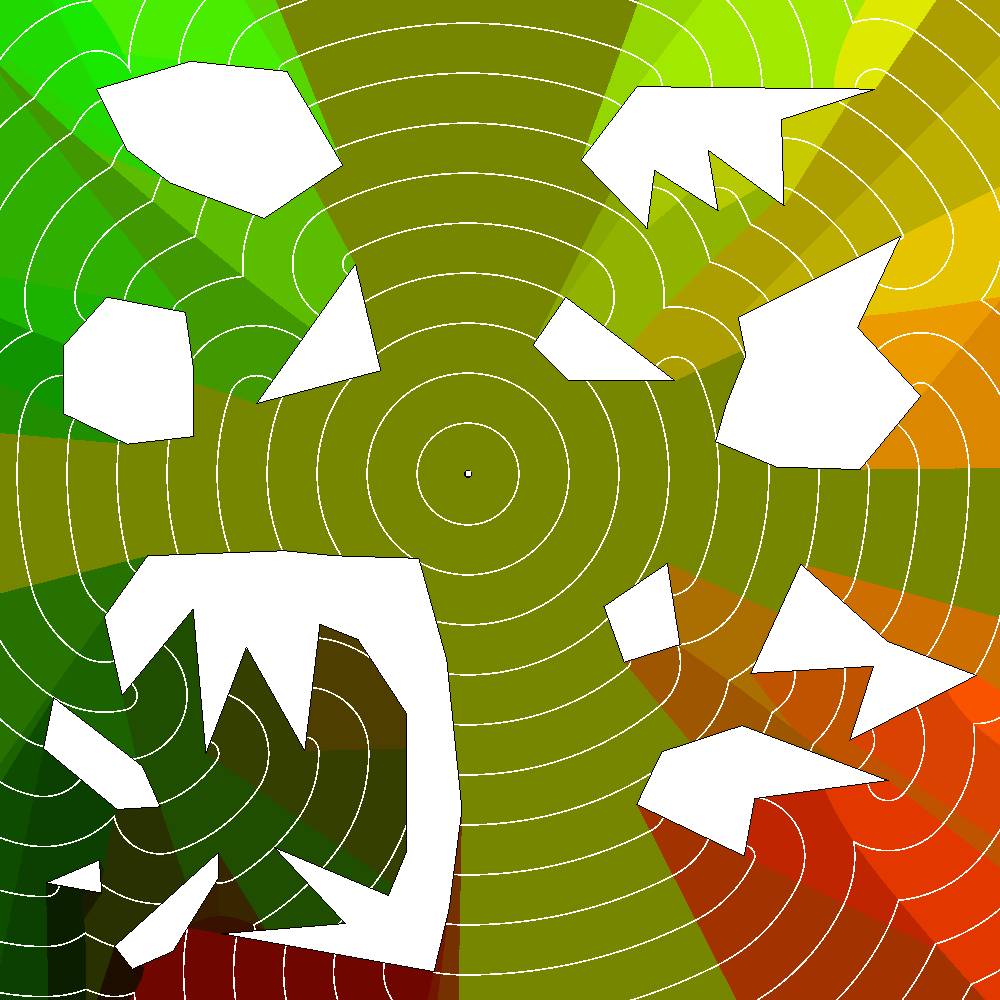}
  \includegraphics[width=0.24\linewidth]{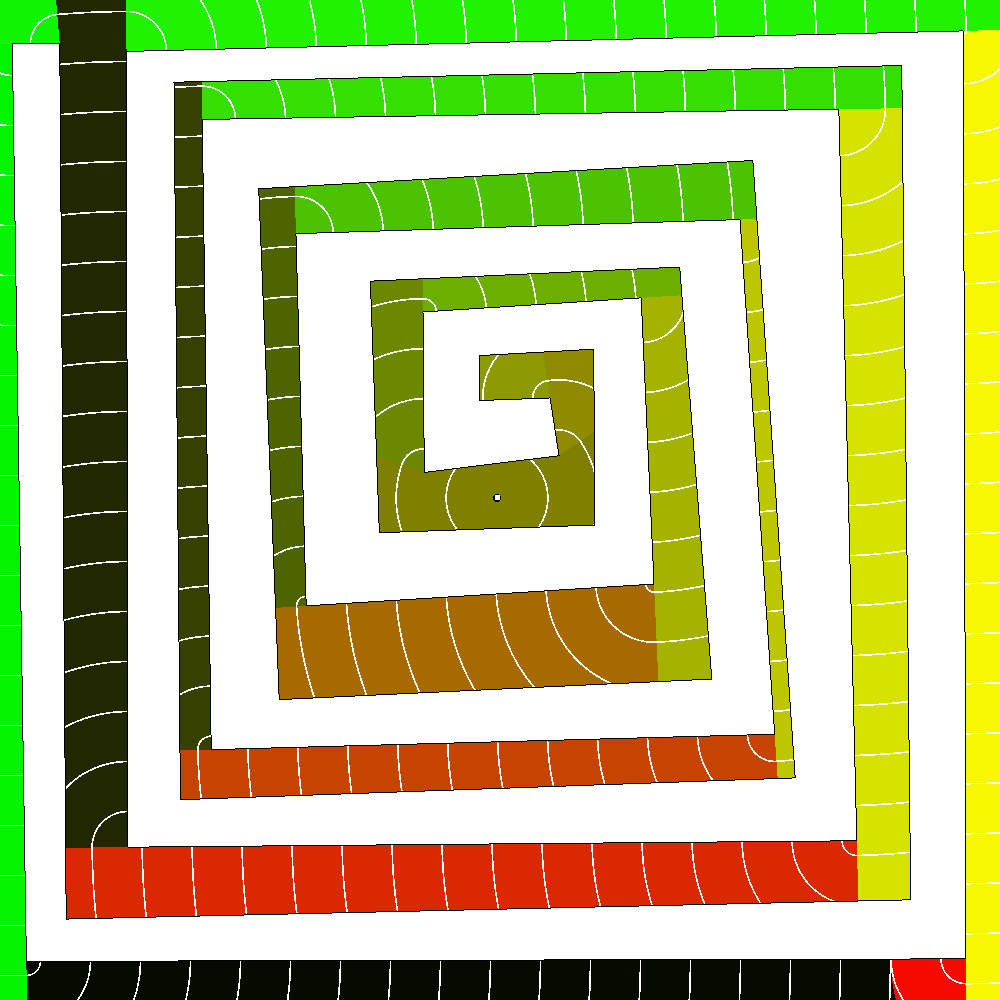}
  \includegraphics[width=0.24\linewidth]{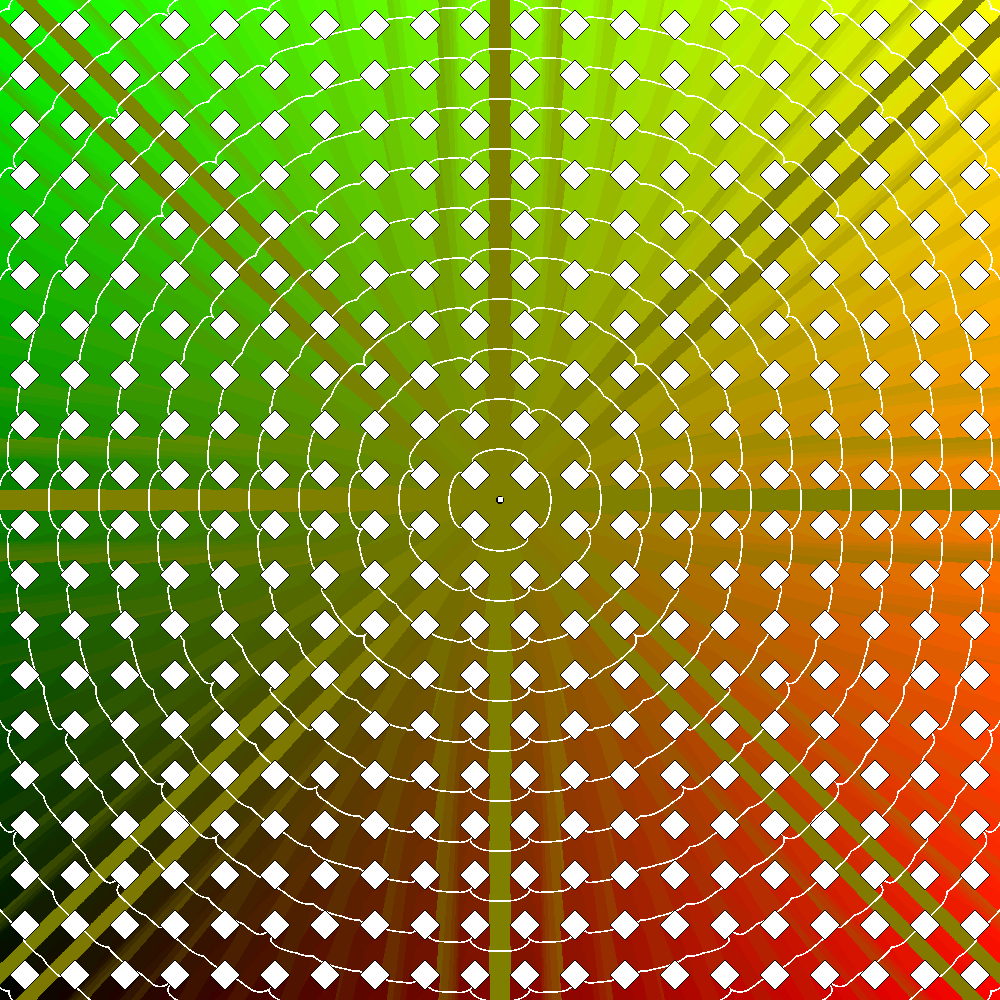}
  \caption{\label{fig:results1}
  Single-source SPM results.
           }
\end{figure*}

\begin{figure*}[htb]
  \centering
  \includegraphics[width=0.24\linewidth]{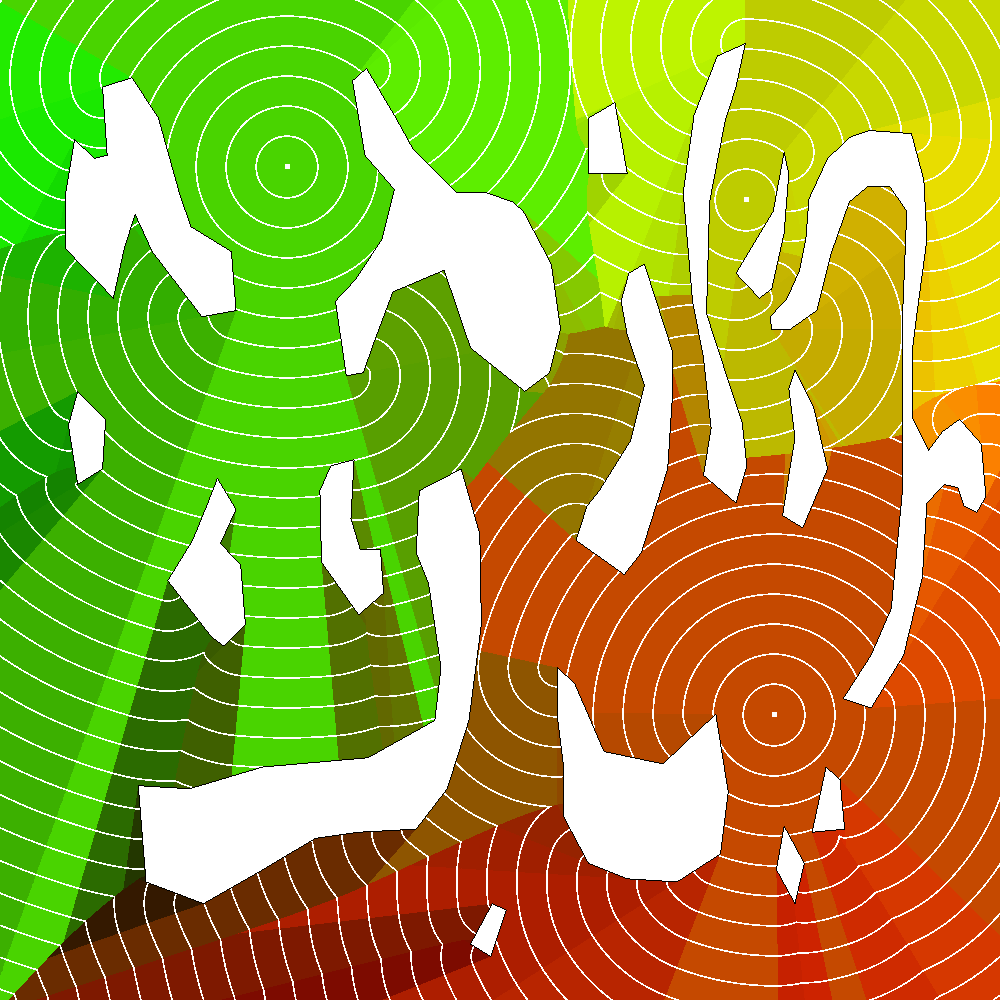}
  \includegraphics[width=0.24\linewidth]{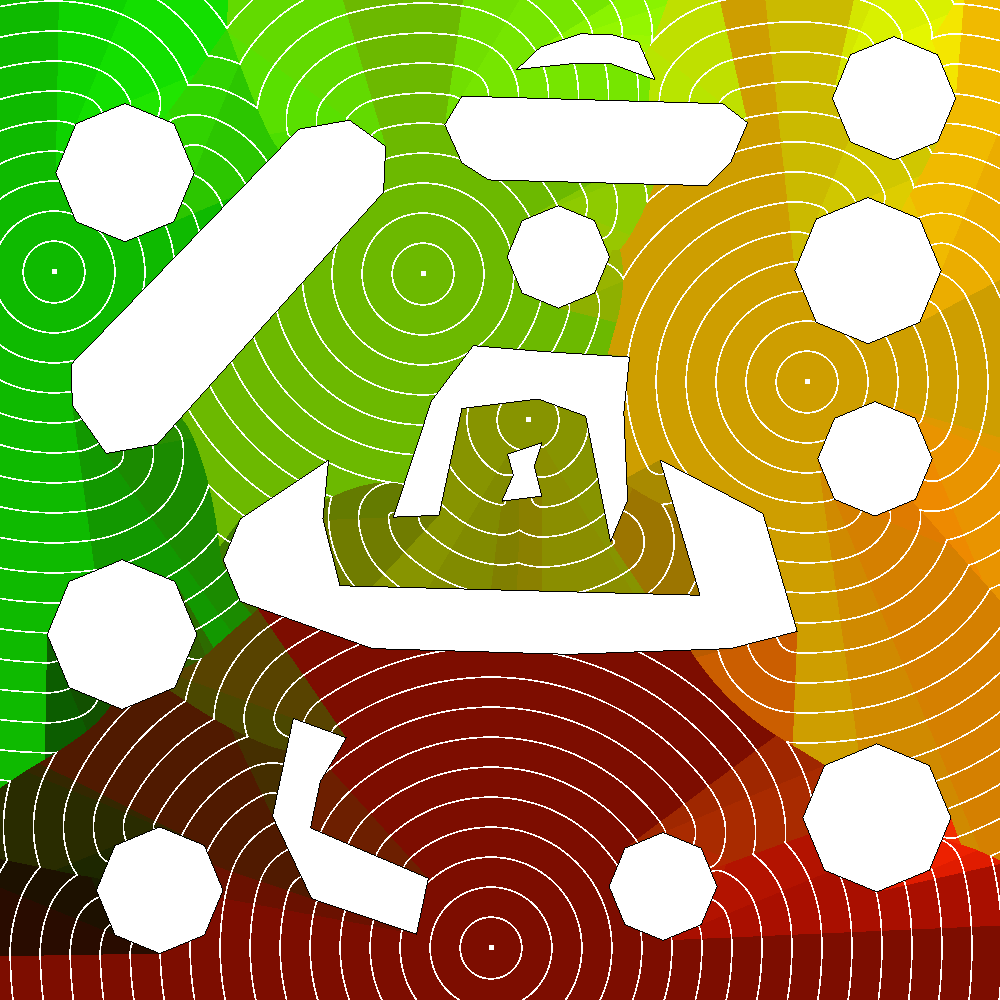}
  \includegraphics[width=0.24\linewidth]{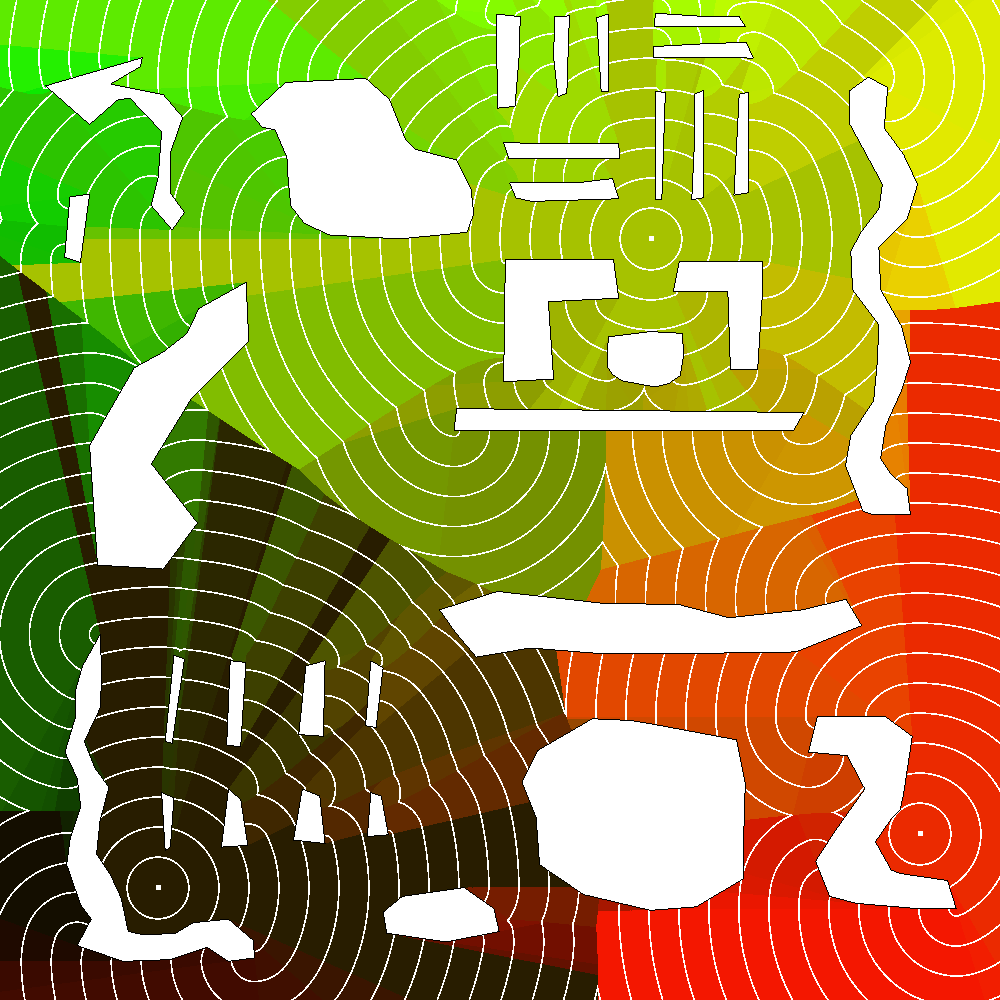}
  \includegraphics[width=0.24\linewidth]{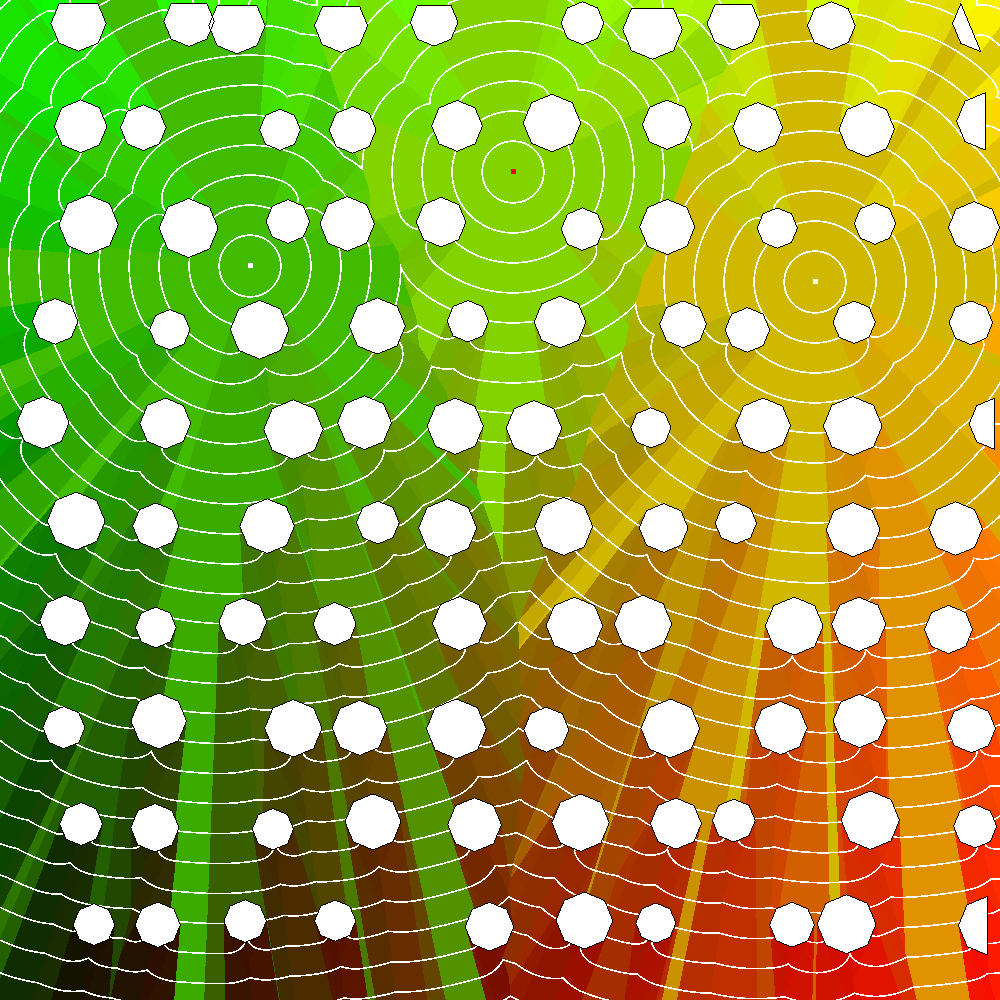}
  \caption{\label{fig:results2}
  Multiple-source SPM results.
           }
\end{figure*}

%% file: 6-discussion.tex
\subsection{Discussion}
\label{sec:discussion}


Although our method uses a framebuffer grid and thus approximates points to the center of the closest pixel when drawing maps, the elimination of discretized cone geometries leads to all distance calculations being computed using the original coordinates of the obstacle vertices. This means that there is no accumulation of error introduced by our method when comparing distances and when integrating the lengths of computed paths.
In practice, when the chosen framebuffer resolution correctly represents an environment, only the region borders formed by collision fronts might be affected by the pixel approximation.
Regions dictate which parent to first take when constructing a shortest path to the closest source, so for a query point that falls on a region border the first vertex choice is subject to a maximum error equal to half a pixel's diagonal.
However even in this case it is possible to eliminate the error by comparing all possible neighboring parent points and choosing the one that is truly the closest to the query point.



Besides being resolution-sensitive the main limitation of our method is that it may only be suitable for real-time simulations in environments of moderate size.
Our method is slower than state-of-the-art path finding solutions that focus on speed of computation instead of global optimality~\cite{Kallmann2014}. 
However, our performance times have potential to increase over time given the rapid expansion of GPU-based computing hardware and techniques.

%% file: 7-conclusions.tex
\section{Conclusions}
\label{sec:conclusions}

We have presented in this paper improved shader-based GPU methods for computing shortest path maps, without the need of pre-computation, and addressing maps with multiple source points and line segments as sources.
These capabilities address practical real-life situations and our benchmarks show that our method outperforms comparable approaches in most cases.

Our approach opens new directions for incorporating navigation information within the traditional graphics pipeline by introducing mapping techniques that can instantly guide agents in multi-agent simulations, with buffers directly storing distances to the closest source and providing the next point to aim for from any point in the environment.

{
\textbf{Acknowledgements}
This research was partially sponsored by the Army Research Office under Grant Number W911NF-17-1-0463. The views and conclusions contained in this document are those of the authors and should not be interpreted as representing the official policies, either expressed or implied, of the Army Research Office or the U.S. Government. The U.S. Government is authorized to reproduce and distribute reprints for Government purposes notwithstanding any copyright notation herein.
The authors also thank Prof. Joseph S. B. Mitchell for several discussions on the topic of this paper.
}